\documentclass[preprint]{aastex}
\usepackage{graphicx}  
\usepackage{fancyheadings}
\usepackage{color}

\begin{document}
\title{Can X-Ray Jets Be Cosmic Beacons?}
%
%
%
%
%
\author{Dan Schwartz\altaffilmark{1},
 }
\altaffiltext{1}{Harvard-Smithsonian Center for Astrophysics}

\begin{abstract}
 If X-rays observed from
\emph{any} extragalactic radio jets are due to inverse Compton
scattering on the cosmic microwave
background (CMB) radiation, then such a source will be detectable with the
same surface brightness anywhere in the more distant universe.
Chandra observations imply that such systems do exist, and will therefore
serve as Cosmic Beacons out to the redshift at which they form.
\end{abstract}

\section{Introduction}
PKS~0637--752, the first celestial X-ray target of the Chandra X-ray
Observatory, revealed a surprisingly strong X-ray jet coincident with
the previously known radio jet \citep{Schwartz00}. The great
difficulty in explaining the X-ray emission \emph{and} allowing the
magnetic field and relativistic electrons to be near equipartition,
led to suggestions~\citep{Cel01,Tav00}
that the extended jet structure was in bulk relativistic motion with
$\Gamma\sim$ 10 on scales greater than 100 kpc. With Chandra detection of X-ray jets soon to be in the tens of
objects, models for at least 3C~273~\citep{Sam01}, and 3C~371~\citep{Pes01} also call for large scale relativistic motions.

\begin{figure}
  \begin{minipage}[t]{0.47\linewidth}
	\centering \includegraphics*[width=2.5in]{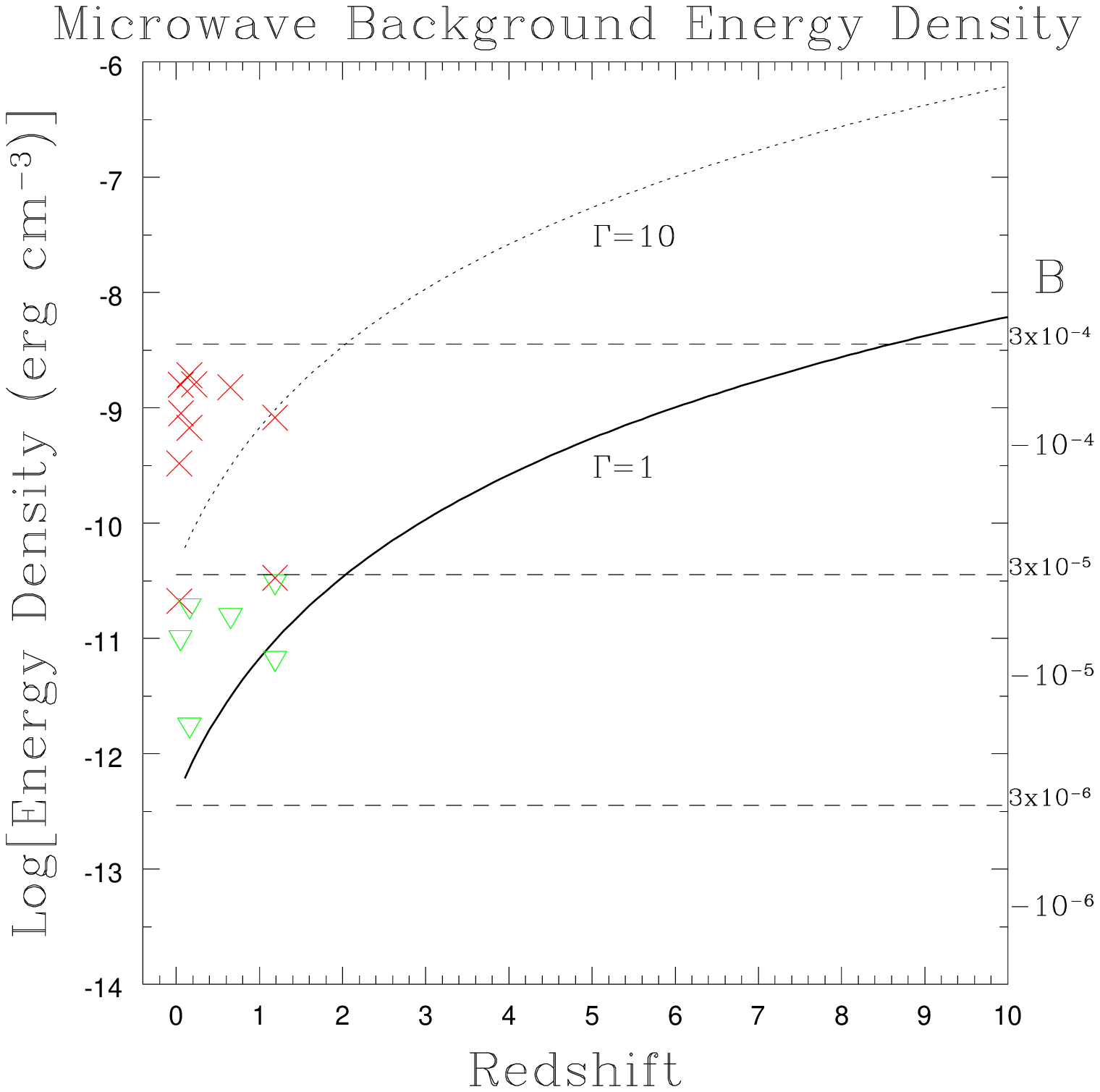} 
 \caption{\label{fig:denvred}
The continuous curves give the apparent CMB energy density (on the left
	vertical scale) vs. redshift as the solid line (in the rest
	frame of the CMB) or the dotted line (in a rest frame moving
	with bulk Lorentz factor $\Gamma=10$ with respect to the
	CMB). The crosses and diamonds plot the redshift and inferred
	magnetic field (on the right vertical scale) for observed
	knots and hotspots in X-ray jets. The crosses assume
	non-relativistic bulk motion, while the diamonds allow the jet
	to have the necessary bulk Lorentz factor to produce the X-ray
	emission while maintaining equipartition in its rest
	frame.  (Data from\protect~\citep{Hard01,Har01,Pes01,Wil00,Wil01})}
  \end{minipage}%
\hspace{0.06\linewidth}
  \begin{minipage}[t]{0.47\linewidth}
	\centering \includegraphics*[width=2.5in]{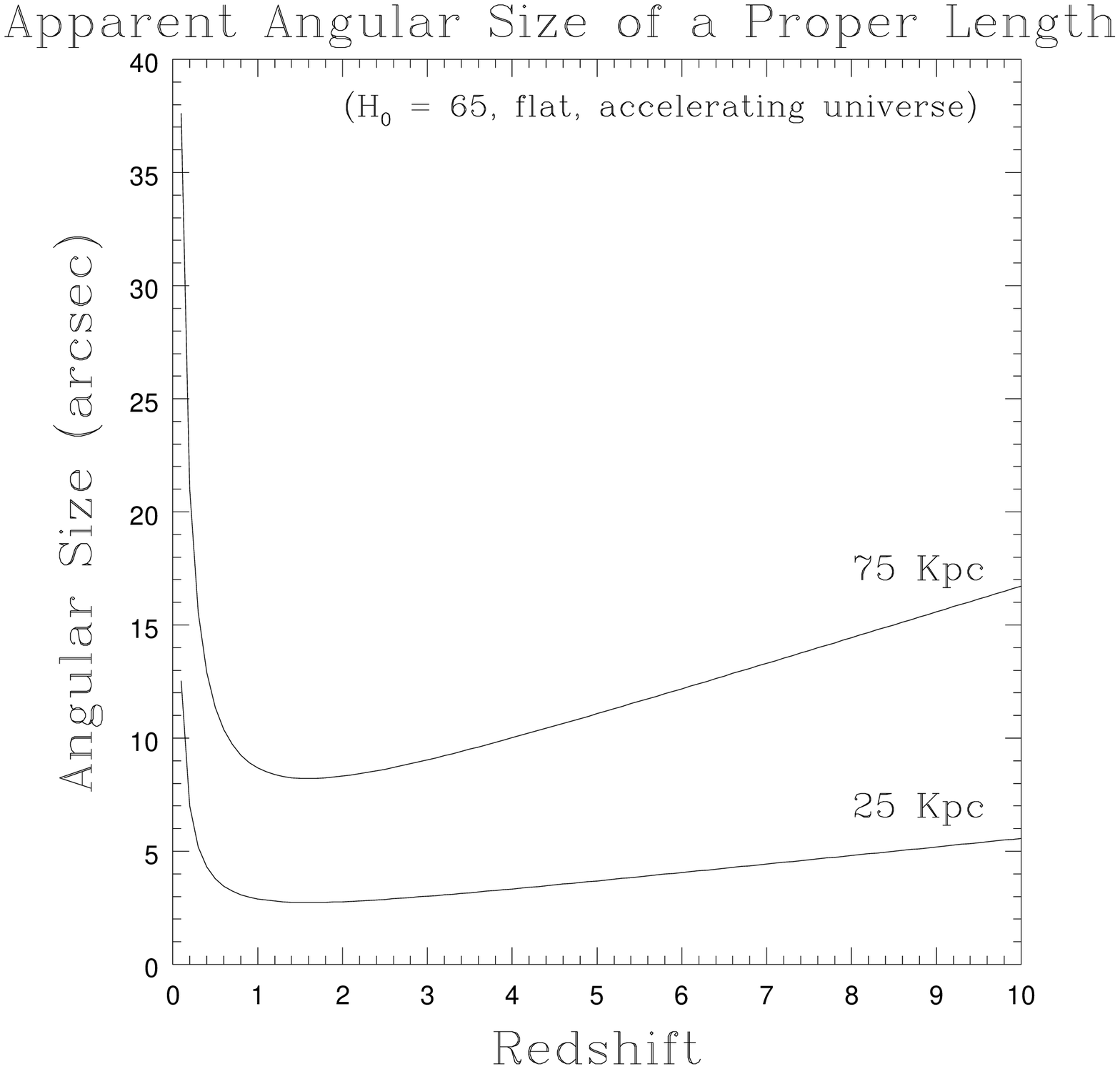}
	\caption{\label{fig:angvred}
The X-ray jets in PKS 0637-752\protect~\citep{Schwartz00}, and 3C 273\protect~\citep{Mar01} brighten about 20 to 30
kpc from the quasar core, and extend 50 to 100 kpc (all distances
projected in the plane of the sky).  We plot the apparent angular size
of projected distances of 25 and 75 kpc as a function of redshift. With
the $1^{\prime\prime}$ resolution of Chandra, 10 kpc can be resolved anywhere in the
universe. Throughout this paper we assume a flat, accelerating
cosmology, with H$_0$=65, $\Omega_0 = 0.3$ , and
$\Omega_{\Lambda}=0.7$, and use the formulas given by \protect~\citet{Pen99}}
  \end{minipage}%
\end{figure}

We point out here that if relativistic beaming is common, we should
see X-ray jets from those sources pointed toward us \emph{anywhere in the
universe}. Even powerful unbeamed sources may be seen at all redshifts, if
their intrinsic magnetic fields are somewhat less than the typical values
$\sim 2\times10^{-4}$. This follows from the standard relation that
the ratio of IC power to synchrotron power is the ratio of the energy
density of target photons to the magnetic field energy density~\citep{Fel66}. The CMB energy density increases with redshift
according to $(1+z)^4$, (Fig.~\ref{fig:denvred}). This factor compensates for the decrease of surface
brightness due to the expanding universe. X-ray jets are resolved by
Chandra and will always be resolved (on-axis) anywhere in the universe
(Fig.~\ref{fig:angvred}). 

As the jets maintain their apparent surface brightness, they can even
outshine the quasar cores.  Such a case might be
speculated for PKS 2215+020, at z=3.572, where ROSAT detects
an unusually high X-ray to optical flux ratio~\citep{Sie98}. Other
cases should occur among the distant radio emitting Sloan
Survey quasars~\citep{Fan01}.

\section{Observational Implications}
	Jets may be displaced by $5^{\prime\prime}$ to
$10^{\prime\prime}$ from the optical object which emits them.  They may therefore be blank
fields within the best location capability of ROSAT, Chandra, and
XMM/Newton. They might also be incorrect identifications of chance
coincidences with optical objects.
     Jets require $\leq\, 1^{\prime\prime}$  angular resolution for
clear identification. If resolved, the jet should point toward the AGN.
      Algorithms to detect extended sources for
telescopes with poorer resolution, e.g. when more than $5^{\prime}$  offaxis in
Chandra, must fit to a linear shape, \emph{not} to a circularly
symmetric shape, for optimum efficiency.

 \section{What Can We Learn?}
        \begin {itemize}
                \item Cosmogony and Jet Physics

\normalfont
\normalsize

                        \begin {itemize}
                \item Discover distant, early activity in the universe.

                \item Verify the common occurrence of relativistic
                beaming on 100 kpc scales. If this occurs, X-ray jets
                from such objects are certain to be seen at any redshift at which they exist.

                \item Is the CMB the limiting dynamical factor in
                        radio lobe and jet formation? The X-rays are
                        produced with electrons of
                        $\gamma\sim1000/\Gamma$, which will have
                        lifetimes decreasing from $\sim  10^8\,\Gamma$
                        to $10^5\,\Gamma$ 
                        years, for redshifts z=1 to 10, where $\Gamma$ is the bulk Lorentz factor
                        of the jet.

                \item Are jets established before the accretion
                        luminosity in the core becomes significant?
                        (In this case there might be no optical
                        counterpart to an X-ray jet.)

                \item Are there radio quiet X-ray jets? This could happen
                        if ``radio'' electrons at higher $\gamma$ have
                        lost their energy to gamma rays, or if larger
                        jet structures can transport upwards of
                        $10^{46}$ erg s$^{-1}$ while $B\leq10^{-5}$ Gauss.

                \item Do jets produce an appreciable component of the
                X-ray background? Although the diffuse X-ray
                background clearly arises from AGN, there are
                significant implications for producing the observed
                spectral shape, and for the mix of absorbed and
                unabsorbed AGN.

                        \end{itemize}

\large
\bfseries
                \item Cosmology
\normalfont
\normalsize

                        \begin {itemize}
                \item Measure the temperature of the cosmic microwave
                background \emph{in situ}.

                \item Verify the decrease of surface brightness
                        according to $(1+z)^{-4}$ (Tolman effect;~\citep{San61,San01}). 
                        \end{itemize}

        \end{itemize}

\section{Conclusions}
We should vigorously search for distant X-ray jets in both wide field
and deep field surveys. Wide field surveys will discover the
relatively rare but bright objects such as PKS 0637-752, if the
explanation of relativistic beaming with $\Gamma\sim10$ occurs in
Nature. If relativistic beaming is not common, then deep surveys will
detect weak X-ray jets from \emph{all} radio jet sources at redshifts
3 to 10, \emph{if} they exist!

\acknowledgments 
 We acknowledge NASA contract 
NAS8-39073 to the Chandra X-ray Center.


\begin{thebibliography}{}
\bibitem[Celotti et al. (2001)]{Cel01} Celotti, A., Ghisellini,
G., and Chiaberge, M. 2001, MNRAS, \textbf{321}, L1

\bibitem[Fan et al. (2001)]{Fan01} Fan, X. et al. 2001, A.J. \textbf{122}, in press

\bibitem[Felten and Morrison (1966)]{Fel66} Felten, J. E. and
Morrison, P. 1966, Ap.J., \textbf{146}, 686

\bibitem[Hardcastle et al. (2001)]{Hard01}Hardcastle, M.J., Birkinshaw, M. and Worrall,
D.M. 2001, MNRAS, \textbf{323}, L17

\bibitem[Harris and Krawczynski (2002)]{Har01} Harris, D. E., and Krawczynski, H. 2002,
Ap.J., \textbf{565}, in press

\bibitem[Marshall et al. (2001)]{Mar01}Marshall, H.L., et al. 2001, Ap.J.,
\textbf{549}, L167

\bibitem[Pen (1999)]{Pen99}Pen, U-L. 1999, Ap.J. Suppl., \textbf{120}, 49

\bibitem[Pesce et al. (2001)]{Pes01} Pesce, J.E., et al.
2001, Ap.J., \textbf{556}, L79


\bibitem[Sambruna et al. (2001)]{Sam01} Sambruna, et al.
2001, Ap.J., \textbf{549}, L161 

\bibitem[Sandage (1961)]{San61}Sandage, A. 1961, Ap.J., \textbf{133}, 355

\bibitem[Sandage and Lubin (2001)]{San01} Sandage, A. and Lubin, L.M. 2001, A.J.,
\textbf{121}, 2271

\bibitem[Schwartz et al. (2000)]{Schwartz00} Schwartz, D. A., et al. 2000,
Ap.J., \textbf{540}, L69

\bibitem[Siebert and Brinkmann (1998)]{Sie98}Siebert, J. and Brinkmann, W. 1998,
Astron. Astrophys., \textbf{333}, 63 

\bibitem[Tavecchio et al. (2000)]{Tav00} Tavecchio, F.,
Maraschi, L., Sambruna, R. M., and Urry, C. M.  2000, Ap.J., \textbf{544}, L23

\bibitem[Wilson et al. (2000)]{Wil00} Wilson, A.S., Young, A.J., and Shopbell,
P.L., 2000, Ap.J., \textbf{544}, L27

\bibitem[Wilson et al. (2001)]{Wil01} Wilson, A.S., Young, A.J., and Shopbell,
P.L., 2001, Ap.J., \textbf{547}, 740

\end{thebibliography}
\end{document}